\documentclass[epj]{svjour}
\usepackage{graphicx}
\begin{document}
\title{Limit current density in 2D metallic granular packings}

\author{S.Dorbolo, M.Ausloos and N.Vandewalle} 

\institute{GRASP, Institut de Physique B5, Universit\'e de Li\`ege, \\ B-4000
Li\`ege, Belgium.}
\date{Received: date / Revised version: date}
%
\abstract{
The electrical properties 2D of packed metallic pentagons have
been studied.  The characterization of such metallic pentagon heaps (like $i-V$ measurements) has been achieved and has allowed to point out two distinct conduction regimes.  They are separated by a transition line along which the system exhibits a memory effect
behavior due to the irreversible improvement of electrical contacts between pentagons (hot spots).  A limit current density has been found. 
} 
\maketitle

\section{Introduction}               

\par Among the wide panel of physical properties exhibited by granular materials, the electrical properties are rather spectacular since they are the origin of radio\-te\-le\-com\-mu\-ni\-ca\-tion.  Let us remind that the first reported antenna has been found by Branly in 1891 \cite{branly}; this antenna was essentially
composed of a glass tube containing metallic filings.  It is not yet understood why a packing of metallic filings becomes a better conductor when an electromagnetic wave hits the system.  In the 19th century, Calzecchi-Onesti has also found that a similar insulator-conductor transition occurs when a high current flows through such a system \cite{onesti}. Thermal analysis \cite{vdb} has shown that the Calzecchi-Onesti transition is due to the soldering of grains.  The authors have observed a strong increase of the temperature along certain electrical pathes.  This indicates the welding of the beads at the contacts.  The pathes being better conducting shunt the remain of the packing and an electric transition occurs when a continuous weakly resistive path is established.  On the other hand, Bonamy \cite{thesisbon} has found that the $i-V$ curve of one contact between two beads is continous : no sharp transition occurs.  Finally, in \cite{apl}, it was shown that the $i-V$ diagram of a lead bead heap is a combinaison of an exponential behavior of the voltage due to slow microsoldering of the beads followed by a sharp transition as soon as a critical current is reached.  This behavior is only obtained in specific conditions : (i) the filings have to be loose packed and (ii) to be a little oxydized at the grain surface.  Those conditions are rather fuzzy and need quantification.  Experiments should be proposed to obtain a complete picture of the electrical behaviors of the granular matter.

\par It is well known that a 3D granular packing is characterized by a complex system of arches along which force lines take place.  A good way for investigating this network is to study how an electrical current $i$ can flow through the system.  Indeed, Hertz \cite{hertz} observed that electrical contacts are enhanced by pressure; electrical currents tends to follow the paths which are created by the lines of mechanical forces.  From this point of view, the electrical paths form an image of the force chains.

\par In a previous paper \cite{apl}, a 3D granular packing has been studied.  The voltage $V$ has been recorded as a function of the injected current intensity $i$.  High memory effect has been pointed out and two regimes, i.e. a conducting and a non-conducting one, have been found to occur depending on the electrical history of the system.  The transition between both regimes is sharp and the so-called
Calzecchi-Onesti transition has been found.  This transition was interpreted in term of the dielectric breakdown of the oxyde layer as in semiconductor devices \cite{mich}.  The voltage $V$ behavior versus the current $i$ was found to be well fitted by an exponential law like 
\begin{equation} 
V=V_0 (1-\exp(-i/i_0)) 
\end{equation} where $V_0$ is the saturation voltage obtained at high injected current and $(V_0 / i_0)$ is the initial resistance of the packing. 

\par Several questions rise from those experiments especially about the origin of the reversibility of the transition.  The microsoldering of beads is surely expected to be an irreversible process, thus a strong memory effect behavior should be highlighted.   

\par This paper concerns further investigation of the $i-V$ diagram of a 2D metallic grains packing.  For that purpose, a simple system has been considered.  It consists in a packing of 1 mm thick planar pentagons for which edges are 10 mm wide  (Fig.1).  Such a 2D system has been studied in ref.\cite{epje,physA} as far as compaction properties are concerned. It has been shown that the formation of arches inside the packing is responsible for the anisotropy of the force lines in granular packings.  The impossibility for pentagons to be packed into in a compact configuration, whence form intrinsic arches, is thus very interesting to study.

\par In the present work, we report the analysis of the voltage behavior during current cycles.  This allows us to describe the $i-V$ plane as supporting two distinct regimes.  Contrary to bulk materials, granular packing is not only represented by one single curve in the $i-V$ plane but by several trajectories depending on the history of the injected currents in the system.  Memory effects have been pointed out.  The points of the $i-V$ plane where such a granular system may exist define an area limited by the so-called irreversibility line and the intrinsic resistance line.  The irreversibility line is a line along which the electrical resistance is weakened when the current increases.

\par In Sect. 2, the experimental setup and conditions are described.  The next sections are devoted to results and interpretations.  Conclusions are drawn in Sect. 5.

\begin{figure}
\includegraphics[width=8.cm]{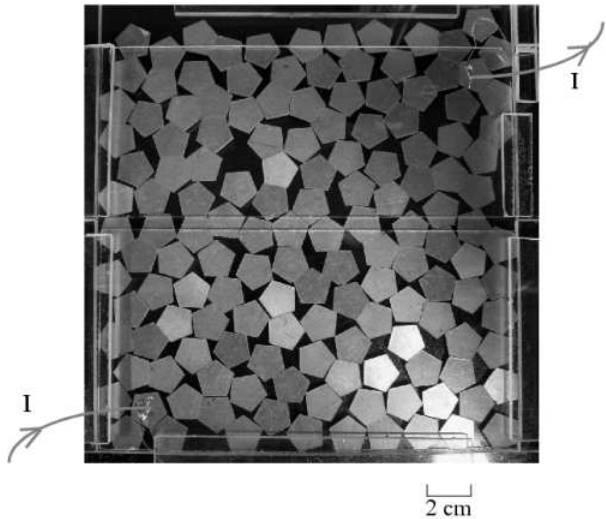}
\caption{Picture of the setup containing the metallic pentagons with current leads.}
\end{figure}

\section{Experimental setup}

About 200 aluminium pentagons are disposed in a plexiglas vertical vessel of $210 \times 200$ mm$^2$.  In order to avoid overlaps between pentagons and, in so doing, to keep 2D conditions, the vessel width is 1 mm thick (Figure 1).  The packing is built by droping the pentagons one by one between vertical planes.  No further compaction process has been performed.  The electrical contacts are set on two particular pentagons.  Those contacts are soldered and the measure of the resistance is performed in a two-wires configuration : the lowest measured global resistance of the packing is about 20 $\Omega$ which is much higher than that of the current leads.

In order to avoid electromagnetic perturbations, the vessel and the measurements instruments are placed in a Faraday cage.  Electrical contacts are connected by coaxial cables to a Keithley K2400 current source which allows to measure the voltage.  During the experiment, we chose to modify the current intensity $i$ and to measure the voltage $V$.

\begin{figure}

\includegraphics[width=8.cm]{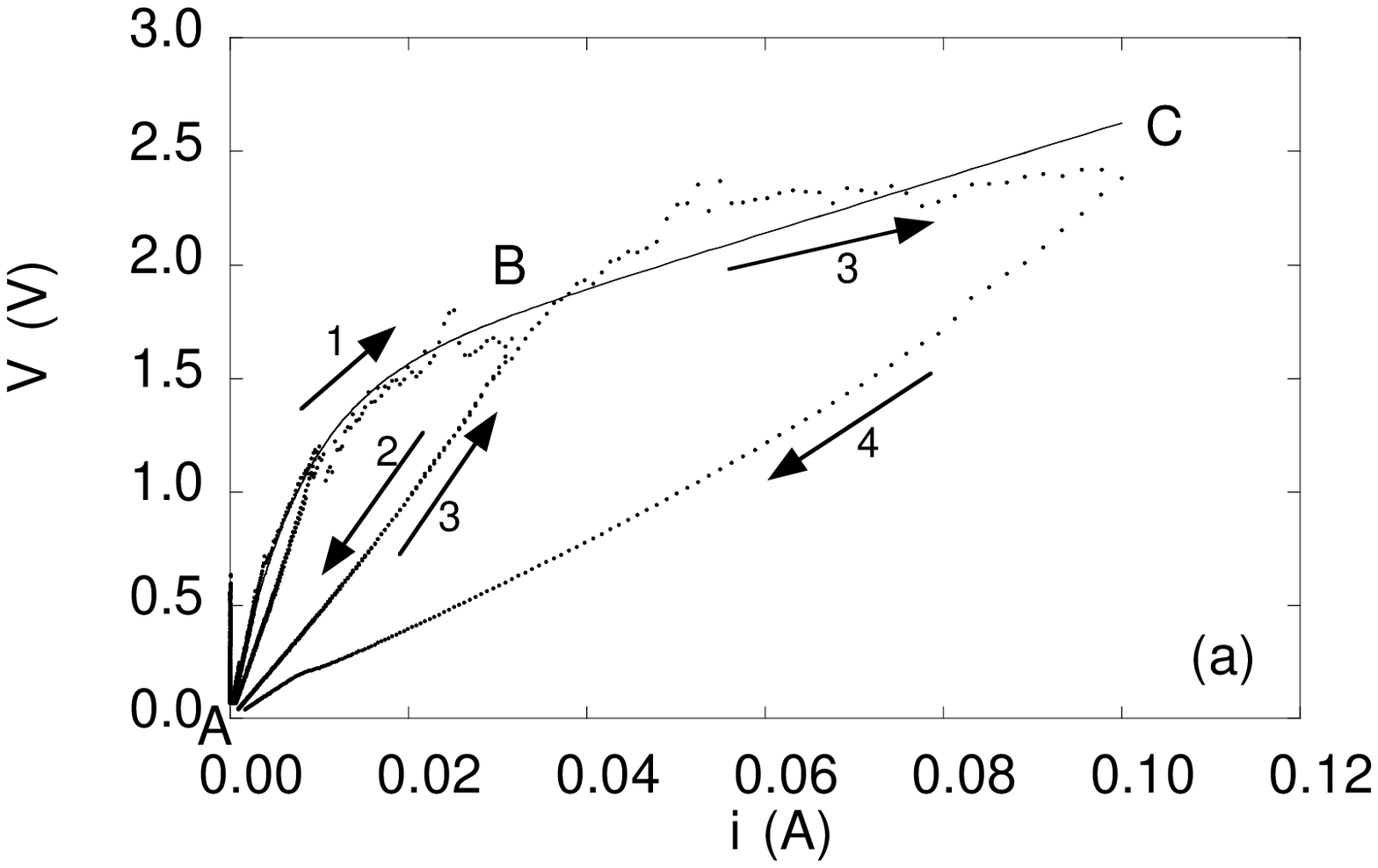}
\includegraphics[width=8.cm]{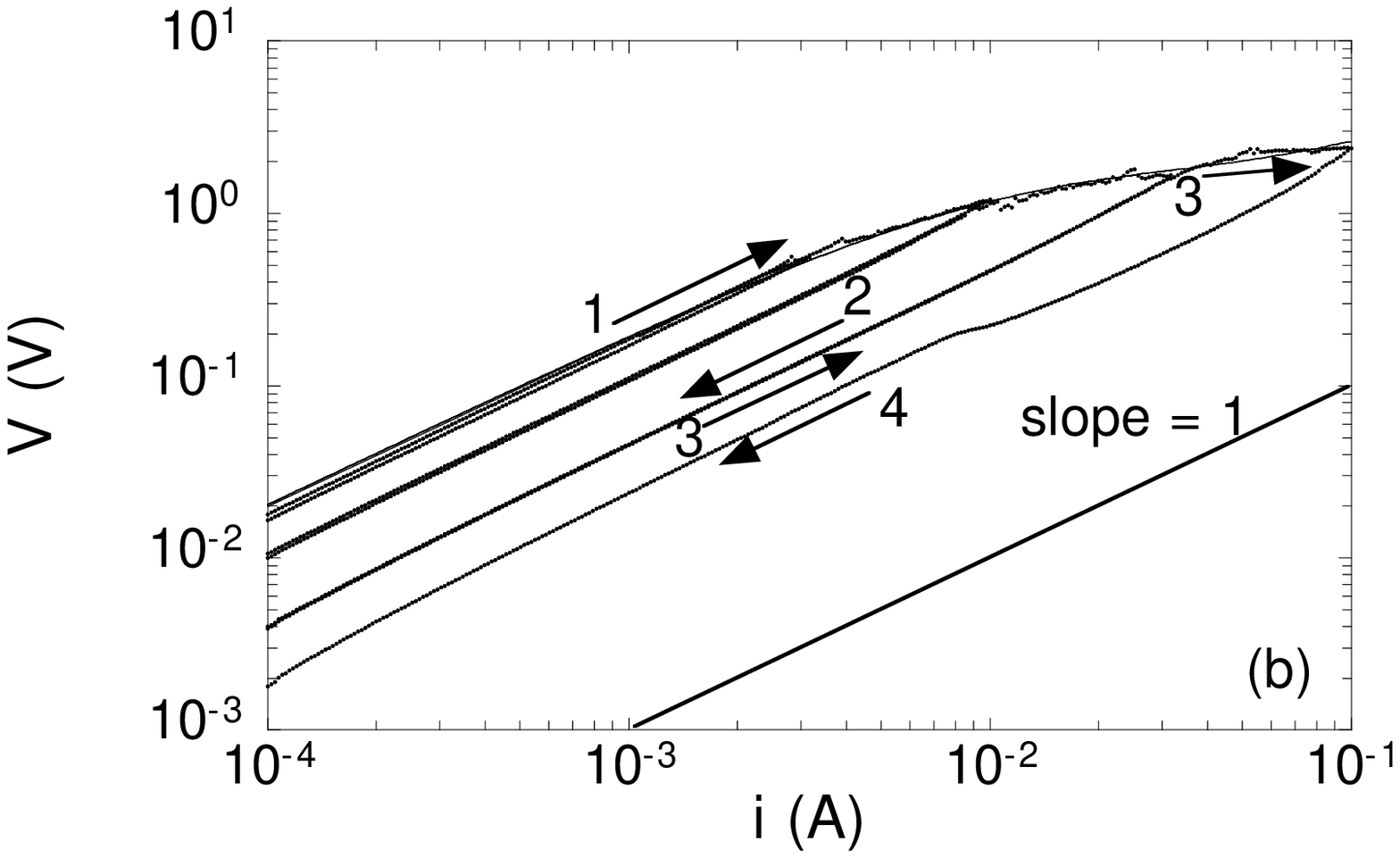}

\caption{(a) The $i-V$ curve obtained for a cycle of injected currents as described in the text, in the upper positive quadrant values. Numbered arrows indicate the different stages of the measurement.  A, B and C are particular points discussed in the text. (b) Same results present in a log-log plot in order to better emphasize the different cycles.} 
\end{figure}

\section{Results} 

The injected
current is set to $i=sign \  10^{-q}$ A,  $q$ being the running parameter in the following description.  A maximum current value is set by fixing $q_{max}$.  Starting from the minimum current $q=6$, the current is increased up to $q=q_{max}$.  It is then decreased to $q=6$ and to ``negative'' values.  To sum up, a cycle is defined as $q$ varies as follow 
\begin{eqnarray} sign=1 : & q = 6  \rightarrow q_{max} \nonumber \\
\/ & q =  q_{max} \rightarrow 6 \nonumber \\
sign=-1 : & q=6  \rightarrow q_{max} \nonumber \\
\/ & q =  q_{max} \rightarrow 6 
\end{eqnarray} such that, a cycle is achieved.  A new value for $q_{max}$ is fixed and another cycle is made.  The maximum values have been successively set to 4, 3, 2.5, 2, 1.5 and
finally 1 which corresponds to the maximum current intensity that the source can deliver (100 mA). In Figure 2a, 2b and 3, the same recorded $i-V$ curves are shown.  Since the range of current intensities is broad, Figure 2b has been represented in a log-log plot.

In order to describe a complete cycle, let us follow the arrows indicated on the Figure 2a, 2b and 3.  Starting from point A, while the current increases, the voltage increases following the law (materialized by a solid line in Figure 2a and 2b) 
\begin{equation} 
V=\left [ V_0 (1-\exp(-i/i_0)) \right ]+ R_r i
\end{equation}
where $R_r$ is the intrinsic electrical resistance of the packing.  The system reaches the point B (10$^{-1.5}$,1.6).  The current is then decreased (arrow 2); the voltage decreases following a linear law $V= R_a \/ i$ where $R_a$ is a fitting coefficient, till the point B' (-10$^{-1.5}$, -1.6) (Figure 3).  Next, the current is increased till 10$^{-1}$ A (arrows 3).  The voltage first follows the same linear law till point B.  After that, it follows the law of Eq.(2) till point C (10$^{-1}$,2.4).  The injected current $i$ is then decreased (arrows 4) : the voltage linearly decreases as $V=R'_a \/i$, where $R'_a$ is lower than $R_a$ and follow the same linear law as the current is increased again to $10^{-6}$ A (arrow 5).

\begin{figure} 
\includegraphics[width=8.cm]{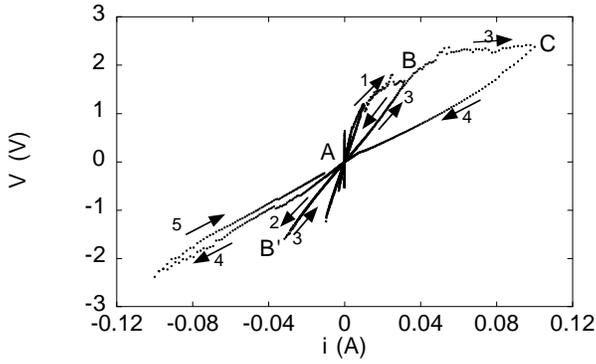}
\caption{The $i-V$ curve obtained during current cycles.}
\end{figure}

A close look at Figure 3 shows that the feature still exists for low injected currents (see Figure 4).  The electrical resistance is very high : about 25 k$\Omega$; while the structure of the $i-V$ curve is similar to Fig.2(a).  This feature would not be observed whether the starting injected current is higher than in this example.  That shows the high hysteretic behavior of such system and that different scales are in presence in a granular packing.  Moreover a very small jump occurs at $i=10^{-4}$ A.  This should be attributed to a breakdown of the reverse diode as observed in a 3D system \cite{apl}.  This breakdown will be explained herebelow.

\begin{figure} 
\includegraphics[width=8.cm]{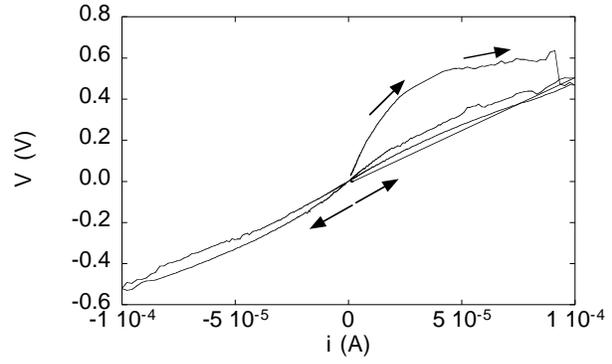}
\caption{Blow-up of the $i-V$ curve of Fig. 3 for very low currents.} 
\end{figure}

\section{Interpretations}

In order to summarize our results, Fig. 5 is a sketch of Figure 2a.  The irreversibility line is described by Eq.(3).  It is thought that microsoldering process occurs along this line since electrical arcs can be seen during the experiment.  The intrinsic resistance line is obtained when the contacts can be neglected.  The resistance of a aluminium wire, representative of the soldered grains, would be then obtained.  Those two lines define in the $i-V$ plane an area included between both lines.  This zone localizes the accessible points by the system.  Ohm's law is of application as indicated by the slope of straight lines in Figure 2b.  Let us consider the point ($i_1$,$v_1$) from this zone.  The straight line which links this point to the origin is given by $V = (v_1 / i_1) i$ (Ohm's law).  This line intersects the curve of Eq.(2) at ($i_c$,$v_2$).   The point ($i_1$,$v_1$) can be thus reached by increasing the current following the irreversibility line to the point ($i_c$,$v_2$) and then by decreasing the current following $V = (v_1 / i_1) i$.  That also means that the Ohm law is applicable in the range of the injected currents [$-i_c$,$i_c$].  In this range, the electrical resistance behaves as a conducting bulk material.

The different behaviors met in the $i-V$ plane find their origin in the  intergrain nature.  A large number of publications concern the modelization of contact network.  They consider either a resistors network \cite{last,kirk,takayasu,dearcangelis,roux,pennetta} or diodes network \cite{hinrichsen,inui} or mixing resistors and diodes network \cite{redner,bigalke}.  Some of those models may reproduce the dielectric breakdown and consider non-linear resistance.  Considering the chemical composition and the geometry of a contact, i.e. an oxyde layer between two metallic pentagons, the non linearity of the electrical resistance can be introduced by considering a granular packing as a network of diodes.  A contact is then represented by two opposite diodes in parallel.  In order to be complete, irreversible processes have to be introduced in the model since memory effect behavior occurs in the system.

We propose to model contacts as if they are made by two opposite diodes in parallel \underline {plus} a resistor $R$ in parallel.  Diodes have a behavior described by Eq.(3).  Resistors are characterized by a section $S$ and a resistivity $\rho$ and follow Ohm's law.  The current will preferencially flow through either the diode or the resistance according to the relative resistance of both components.

\begin{figure} 
\includegraphics[width=8.cm]{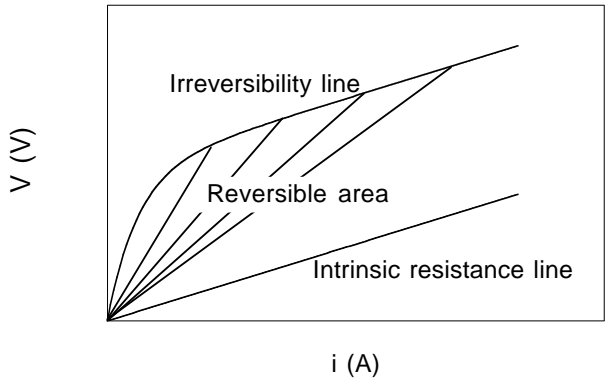}
\caption{Sketch of the different zones and curves presented in the $i-V$ diagram.} 
\end{figure}

Consider the system as being a mean-field media characterised by an average resistivity $\rho$.  The resistance $R$ of the contact can be written using Pouillet law as $R=\rho
\ell/S$ where $\ell$ is the effective length of an electrical path between the electrodes.  When $V=V_0$ at the saturation, it can be rewritten according to Pouillet law : $V_0=\rho \ell i /S$.  The ratio $i/S$ equals to the conductivity $\sigma$ multiplied by the limit electrical field $E_0=V_0/\ell$ (about 5 V/m).  In other words, $i/S$ equals to $J_{\ell}$
which is hereby called a limit current density.  This quantity remains constant in the range of high currents.  This means that an increase of current intensity corresponds to an increase of the section of the contact resistor so as to keep $J_{\ell}$ constant.  The grains knit together following the rule $S=J_{\ell} /i$.  Within this approximation, the section increases of a factor 3 between the point B and C in Fig.2a.  This microsoldering process is of course irreversible.  When the current intensity decreases, the current follows the path defined by those enhanced contacts which shunts the diodes.  The Ohm's law is from then on application as shown in Fig. 2a and 5.  

The very low current feature shown in Fig.4 exhibits rather the same behavior as described here above.  When the same arguments are applied to this, a larger $J_{\ell}$ is found since $E_0=2.5$ V/m in this case.  Nevertheless, a breakdown occurs at 10$^{-4}$ A (Fig.4).  This is
similar to the one in a 3D system as in ref.\cite{apl}.  It can be explained as a Zener or avalanche breakdown of the reverse diode when the voltage across a contact reaches the critical value, i.e. the Calzecchi-Onesti transition.

\section{Conclusions}

The above measurements show that a weak electrical contact in granular metallic packings has to be modelized as three
components in parallel : two opposite diodes and a resistor.  This resistor modelizes the microsoldering and is characterized by a limit current density.  When this limit density is reached, the equivalent section of the resistor increases, the grains
weld together.  This is an irreversible process.  The model allows to explain
the hysteretic loops in $i-V$ curves.  Moreover electrical properties of metallic grains packing are strongly determined by the electrical history of the system. Fine structures can be pointed out or not according to injected current cycles.

\vskip -0.25cm
\section*{Acknowledgements} S.D. is a F.N.R.S Scientific Research Worker.  We would like to thank Prof. H.W. Vanderschueren for the
use of MIEL
facilities. We want also to thank Dr. M. Houssa (U. Marseille 3) and Dr. R. Goffaux
(SEO, Luxemburg) for comments.

\end{document}